\documentclass[aps, prd, letterpaper, twocolumn, amsmath, amssymb, preprintnumbers, superscriptaddress, floatfix, nofootinbib, longbibliography]{revtex4-2}
\usepackage[T1]{fontenc}
\usepackage{graphicx}
\usepackage{amsmath}
\usepackage{amsfonts}
\usepackage[colorlinks=true, linkcolor=blue, urlcolor=blue, citecolor=blue, anchorcolor=blue]{hyperref}
\usepackage{bbold}
\usepackage{tikz-feynman}
\tikzfeynmanset{compat=1.0.0}

\newcommand{\ignore}[1]{}

\newcommand{\cL}{\ensuremath{\mathcal L} }

\newcommand{\Tr}[1]{\ensuremath{\mbox{Tr}\left[ #1 \right]} }

\newcommand{\ket}[1]{\left| #1 \right>} % for Dirac bras
\newcommand{\matrixel}[3]{\left< #1 \vphantom{#2#3} \right|
	#2 \left| #3 \vphantom{#1#2} \right>} % for Dirac matrix elements
\begin{document}
\preprint{RIKEN-iTHEMS-Report-23}
\preprint{LLNL-JRNL-853554}
\preprint{FERMILAB-CONF-23-260-T}
%\preprint{SI-HEP-2021-18}
\title{Hidden Conformal Symmetry from the Lattice}

\author{T.~Appelquist}
\affiliation{Department of Physics, Sloane Laboratory, Yale University, New Haven, Connecticut 06520, USA}
\author{R.~C.~Brower}
\affiliation{Department of Physics and Center for Computational Science, Boston University, Boston, Massachusetts 02215, USA}
\author{K.~K.~Cushman}
\affiliation{Department of Physics, Sloane Laboratory, Yale University, New Haven, Connecticut 06520, USA}
\author{G.~T.~Fleming}
\affiliation{Department of Physics, Sloane Laboratory, Yale University, New Haven, Connecticut 06520, USA}
\affiliation{Theoretical Physics Division, Fermilab, Batavia, IL 60510, USA}
\author{A.~Gasbarro}
\affiliation{AEC Institute for Theoretical Physics, University of Bern, 3012 Bern, CH}
\author{A.~Hasenfratz}
\affiliation{Department of Physics, University of Colorado, Boulder, Colorado 80309, USA}
\author{J.~Ingoldby}\email{ingoldby@ictp.it}
\affiliation{Abdus Salam International Centre for Theoretical Physics, Strada Costiera 11, 34151, Trieste, Italy}
\author{X.~Y.~Jin}
\affiliation{Computational Science Division, Argonne National Laboratory, Argonne, Illinois 60439, USA}
%\author{J.~Kiskis}
%\affiliation{Department of Physics, University of California, Davis, California 95616, USA}
\author{E.~T.~Neil}
\affiliation{Department of Physics, University of Colorado, Boulder, Colorado 80309, USA}
\author{J.~C.~Osborn}
\affiliation{Computational Science Division, Argonne National Laboratory, Argonne, Illinois 60439, USA}
\author{C.~Rebbi}
\affiliation{Department of Physics and Center for Computational Science, Boston University, Boston, Massachusetts 02215, USA}
\author{E.~Rinaldi}
\affiliation{Interdisciplinary Theoretical and Mathematical Sciences Program (iTHEMS), RIKEN, 2-1 Hirosawa, Wako, Saitama 351-0198, Japan}
\author{D.~Schaich}
\affiliation{Department of Mathematical Sciences, University of Liverpool, Liverpool L69 7ZL, UK}
\author{P.~Vranas}
\affiliation{Physical and Life Sciences Division, Lawrence Livermore National Laboratory, Livermore, California 94550, USA}
\affiliation{Nuclear Science Division, Lawrence Berkeley National Laboratory, Berkeley, California 94720, USA}
\author{E.~Weinberg}
\affiliation{Department of Physics and Center for Computational Science, Boston University, Boston, Massachusetts 02215, USA}
\affiliation{NVIDIA Corporation, Santa Clara, California 95050, USA}
\author{O.~Witzel}
\affiliation{Center for Particle Physics Siegen (CPPS), Theoretische Physik 1, Naturwissenschaftlich-Technische Fakult\"at, Universit\"at Siegen, 57068 Siegen, Germany}

\collaboration{Lattice Strong Dynamics (LSD) Collaboration}
\noaffiliation

%\date{\today}
% ------------------------------------------------------------------

% ------------------------------------------------------------------
\begin{abstract}
We analyze newly expanded and refined data from lattice studies of an $SU(3)$ gauge theory with eight Dirac fermions in the fundamental representation. We focus on the light composite states emerging from these studies, consisting of a set of pseudoscalars and a single light scalar. We first consider the view that this theory is just outside the conformal window. In this case, the pseudoscalars arise from spontaneous breaking of chiral symmetry. Identifying the scalar in this case as an approximate dilaton, we fit the lattice data to a dilaton effective field theory, finding that it yields a good fit even at lowest order. For comparison, we then consider the possibility that the theory is inside the conformal window. The fermion mass provides a deformation, triggering confinement. We employ simple scaling laws to fit the lattice data, and find that it is of lesser quality.
\end{abstract}

\maketitle
% ------------------------------------------------------------------

% ------------------------------------------------------------------
\section{Introduction}
\label{sec:intro}

The study of the Higgs boson at the LHC shows it to be well described by the standard model \cite{ATLAS:2012yve,CMS:2012qbp,ATLAS:2020qdt,CMS:2020gsy}. This description could, however, evolve into something more fundamental at higher energy scales, one possibility being compositeness emerging from underlying strong interactions. The relative lightness of the Higgs boson could then be accounted for if it is a pseudo-Nambu-Goldstone boson (pNGB) generated by the spontaneous breaking of an approximate internal global symmetry~\cite{Kaplan:1983fs,Panico:2015jxa,Cacciapaglia:2020kgq,Banerjee:2022xmu} or a conformal symmetry, the latter leading to a description of the Higgs boson as an approximate dilaton~\cite{Goldberger:2007zk}. In recent studies~\cite{BuarqueFranzosi:2018eaj,Appelquist:2020bqj}, a composite Higgs boson is interpreted as an admixture of pNGB and dilaton.

Important elucidation of these ideas has come from the study of a class of lattice $SU(N)$ gauge field theories~\cite{DeGrand:2015zxa,Svetitsky:2017xqk,Witzel:2019jbe,Drach:2020qpj}. With the number $N_f$ of light Dirac fermions taken to be relatively large compared to the number of colors, these theories can be near or within the ``conformal window'' where confinement gives way to infrared conformality. For the gauge group $SU(3)$, the smallest such value is thought to be around $9$, but the exact value is not known~\cite{Ryttov:2016hal,Ryttov:2017kmx,Antipin:2018asc,DiPietro:2020jne}. 

Lattice studies of the $N_f = 8$ theory~\cite{Appelquist:2007hu,Deuzeman:2008sc,Fodor:2009wk,Hasenfratz:2014rna,Fodor:2015baa,Aoki:2016wnc,Appelquist:2016viq,Appelquist:2018yqe,Kotov:2021mgp,Hasenfratz:2022zsa,Hasenfratz:2022qan,LSD:2023aaa} have suggested that it is close to the conformal window boundary. If it is outside and confining, the light pseudoscalars in the spectrum are pNGBs of a spontaneously broken internal symmetry. The fermion masses employed in the lattice calculations give them a small mass. The spectrum also includes a scalar particle that is lighter than the vector meson, and also lighter than its analog in $N_f=2$ lattice QCD, once larger unphysical quark masses are used so that the pNGB to vector meson mass ratios in the two theories are similar. See \cite{LSD:2023aaa,Rodas:2023twk} and references therein.

The lightness of the scalar is possibly attributable to the theory's nearness to the conformal window. It could be an approximate dilaton - the pseudo-Goldstone boson that emerges when the breaking of scale invariance is dominantly spontaneous. A dilaton-effective-field-theory (dEFT) description of these states~\cite{Golterman:2016lsd,Golterman:2016hlz,Kasai:2016ifi,Fodor:2017nlp,Golterman:2018mfm,Fodor:2020niv,Golterman:2020utm,Golterman:2021ohm,Appelquist:2022mjb} has been found to fit earlier $N_f = 8$ lattice datasets well~\cite{Appelquist:2017wcg,Appelquist:2019lgk,Fodor:2019vmw,Golterman:2020tdq,LatticeStrongDynamicsLSD:2021gmp}.

An alternate possibility is that the $N_f = 8$ theory is inside the conformal window. The fermion mass would then provide a deformation of the infrared conformal symmetry, with confinement triggered at a comparable scale. Simple scaling laws describe the masses and decay constants of all the composite states~\cite{DelDebbio:2010ze,DelDebbio:2010jy}, which extrapolate to zero in the zero-fermion-mass limit. There is no approximate conformal or internal global symmetry to be spontaneously broken. There would be no evident EFT, the scaling laws being described directly in terms of the underlying gauge theory.

In this letter, we reexamine the light spectrum of the $SU(3)$ gauge theory with $N_f = 8$, drawing on our latest lattice data. We first consider the view that it is outside the conformal window. The dEFT provides a good fit to the lattice data, now incorporating data for a scalar decay constant $F_S$. We also consider the possibility that the theory is inside the conformal window, described in the infrared by mass-deformed conformal symmetry. We find that this description is less consistent with the lattice data than dEFT.

% ------------------------------------------------------------------
\section{Summary of Lattice Results}
\label{sec:latt}

We draw on lattice data presented in a companion paper~\cite{LSD:2023aaa} for the pseudoscalar mass $M_\pi$, the pseudoscalar decay constant $F_\pi$, the mass of the light scalar $M_d$ (identified with the dilaton in dEFT), and a scalar decay constant $F_S$, in units of the lattice spacing. A common mass is given to all eight flavors, and we compute data for five different values of this fermion mass.

Our numerical data are obtained using improved nHYP–smeared staggered
fermions with parameters $\alpha \equiv (0.5, 0.5, 0.4)$ and a gauge
action containing both fundamental and adjoint plaquette terms with coefficients $\beta_F = 4.8$ and $\beta_A = -\beta_F /4$
for all data points. The dataset in Ref.~\cite{LSD:2023aaa} extends our previously generated $N_f=8$ data \cite{Appelquist:2016viq,Appelquist:2018yqe} and uses improved analysis techniques. Compared to the earlier studies, we include infinite volume extrapolations and improved estimates of other systematic errors. The data for meson masses and decay constants that we use in the present study are summarized in Table IX of Ref.~\cite{LSD:2023aaa}. We find that the statistical covariances between different observables are small and we neglect them here. The relevant lattice ensembles are shown in Tables XI and XII.

We also include data for $2\rightarrow 2$ pNGB scattering presented in Ref.~\cite{LatticeStrongDynamicsLSD:2021gmp}. These are for $k\cot\delta(k)$ in the s--wave, $I = 2$ isospin channel. On the largest lattice volumes shown there, the scattering momentum $k$ is very small and can be neglected. We therefore identify these data points with the inverse scattering length $1/a^{(2)}_{0}$, expressed in units of the lattice spacing.

Since we consider two interpretations of the lattice data, we employ the Akaike Information Criterion (AIC)~\cite{Akaike:1974AIC} to help assess which provides the better description. It is given by
\begin{equation}
	\text{AIC} \equiv \chi_{\text{min}}^2 + 2k \,,
	\label{eq:aic}
\end{equation}
where $k$ is the number of free parameters in the model. The term $\chi^2_{\text{min}}$ is the standard chi-squared function for the fit to data, minimized with respect to the model parameters. Models having a smaller AIC value may be interpreted as more probable \cite{Jay:2020jkz} in a Bayesian framework. While differences between AIC values are meaningful, the absolute value of the AIC does not on its own give information about the fit quality.

% ------------------------------------------------------------------
\section{Dilaton EFT}
\label{sec:dilaton}

To describe the spontaneous breaking of approximate dilatation symmetry, we introduce a scalar dilaton field $\chi$, which acquires a nonzero vacuum value $\langle\chi\rangle=f_d$. There is also an approximate global symmetry $SU(N_f)\times SU(N_f)$ that is spontaneously broken to a subgroup $SU(N_f)$, so we introduce a multiplet of pNGB fields $\Sigma$ satisfying the constraint $\Sigma^\dagger\Sigma=\mathbf{1}$.

The EFT Lagrangian, reviewed in Ref.~\cite{Appelquist:2022mjb}, is given by
\begin{multline}
	\cL = \frac{1}{2}\partial_{\mu}\chi\partial^{\mu}\chi \, + \frac{f_{\pi}^2}{4}\left(\frac{\chi}{f_d}\right)^2 \, \Tr{\partial_{\mu}\Sigma(\partial^{\mu}\Sigma)^{\dagger}}\\ + \frac{m_{\pi}^2 f_{\pi}^2}{4} \left(\frac{\chi}{f_d}\right)^y \, \Tr{\Sigma + \Sigma^{\dagger}} \, - \, V_{\Delta}(\chi) \, ,
	\label{eq:L}
\end{multline}
where we include only the leading order (LO) terms in a low energy expansion. We identify $m^2_\pi= 2B_\pi m$, where $m$ is the fermion mass in the underlying gauge theory and $B_\pi$ is a constant. The exponent $y$ is taken as a free parameter in fits to lattice data.

We take the scalar potential to be \cite{Goldberger:2007zk}
\begin{equation}
	V_{\Delta}(\chi) = \frac{m_d^2\chi^4}{4(4-\Delta)f_d^2}\left[1-\frac{4}{\Delta}\left(\frac{\chi}{f_d}\right)^{\Delta-4} \right] \, ,
	\label{eq:vdelta}
\end{equation}
where $\Delta$ is a free parameter determined from the lattice data. In the limit $\Delta\rightarrow 4$, this potential takes a logarithmic form.

With $m^2_\pi$ nonzero, $\chi$ develops a new vacuum value $\langle\chi\rangle=F_d$. By minimizing the potential terms for the dilaton in Eqs.~(\ref{eq:L}) and (\ref{eq:vdelta}), we find
\begin{equation}
	\frac{F_d^{4-y}}{(4-\Delta)f^{4-y}_d}\left[1-\left(\frac{f_d}{F_d}\right)^{4-\Delta}\right]=\frac{yN_ff^2_\pi m^2_\pi}{2f^2_dm^2_d}.\label{eq:min}
\end{equation}
By normalizing the kinetic term for the pNGB field in the new vacuum, we arrive at the scaling relations
\begin{align}
	\frac{F_{\pi}^2}{f_{\pi}^2} = \frac{F_{d}^2}{f_{d}^2} \, ,\qquad\qquad
	\frac{M_{\pi}^2}{m_{\pi}^2} = \left( \frac{F_{d}}{f_{d}}\right)^{y-2} \, . \label{eq:mpscaling}
\end{align}

The physical dilaton mass is given by
\begin{equation}
	\label{eq:mdscaling}
	\frac{M^2_d}{F^2_d} = \frac{m^2_d}{(4-\Delta)f^2_d}\left(4-y+(y-\Delta)\left(\frac{f_d}{F_d}\right)^{4-\Delta}\right).
\end{equation}

The dEFT prediction \cite{LatticeStrongDynamicsLSD:2021gmp} for the s-wave scattering length for two pNGBs in the isospin-2 channel is given by
\begin{align}
	M_\pi a^{(2)}_{0} = -\frac{M^2_\pi}{16\pi F^2_\pi}\left(1-(y-2)^2\frac{f^2_\pi}{f^2_d}\frac{M^2_\pi}{M^2_d}\right)\,.\label{eq:dilatonslength}
\end{align}
The first term in the large parentheses would be present if there was no light dilaton in the spectrum~\cite{Weinberg:1966kf}. The second term comes from the exchange of a virtual dilaton.

We include an additional quantity that we have recently measured on the lattice: the scalar decay constant $F_S$ defined in terms of the matrix element
\begin{equation}
	\matrixel{0}{J_{S}(x)}{\chi(p)} \equiv F_{S} M_d^{2}e^{-ip\cdot x}\,,
\end{equation}
where $\ket{\chi(p)}$ refers to the dilaton state and $J_{S}(x) = m \bar{\psi}\psi$ is the scheme independent scalar current~\cite{Gasser:1983yg}. The quantity $F_S$ may be extracted from the dilaton contribution to the two point correlator $\langle J_S(x)J_S(0)\rangle$. The technique is demonstrated for the chiral Lagrangian EFT in Refs.~\cite{Gasser:1983yg,Pich:2018ltt}. We find
\begin{equation}
	|F_S| = \frac{yN_fM^2_\pi F_\pi}{2M^2_d}\frac{f_\pi}{f_d}.
	\label{eq:dilfspred}
\end{equation}
A related formula for $F_S$ was derived using current algebra and compared with lattice data in Ref.~\cite{LatKmi:2015non}.

The quantity $F_S$ would control the decay rate of the dilaton if the gauge theory were modified by including a heavy scalar mediator with a weak Yukawa coupling to the fermion bilinear and to other light states. The mediator would then enable the decay of the dilaton to these states. Both $F_S$ and this decay rate would vanish in the chiral limit since the scalar current $m \bar{\psi} \psi$ defining $F_S$ and the Yukawa interaction that enables the decay both break chiral symmetry.

We perform a global fit at leading order to the lattice data
for $M_\pi$, $F_\pi$, $M_d$, $F_S$ and $M_\pi\,a^{(2)}_0$. This is the first
time that lattice data for $F_S$ has been used in an EFT fit. The fit incorporates $N = 25$ lattice data points and $k = 6$ free model parameters, which we take to be $\{y,\,B_\pi,\,\Delta,\,f^2_\pi,\,f^2_\pi/f^2_d,\,m^2_d/f^2_d\}$. We obtain dEFT predictions for the quantities measured on the lattice by first solving Eq.~(\ref{eq:min}) to obtain $F_d/f_d$. We determine $M_\pi$ and $F_\pi$ from the scaling relations shown in Eq.~(\ref{eq:mpscaling}), and we use Eqs.~(\ref{eq:mdscaling}), (\ref{eq:dilatonslength}) and (\ref{eq:dilfspred}) to determine the remaining quantities.

\begin{table}
	\centering
	\vspace{12pt}
	\renewcommand\arraystretch{1.2}
	\begin{tabular}{| c || c | c |}
		\hline
		Parameter & LO & NLO\\
		\hline\hline
		$y$ & 2.091(32) & 2.069(32)\\
		$B_\pi$ & 2.45(13) & 2.46(13)\\
		$\Delta$ & 3.06(41) &2.88(49)\\
		$f^2_\pi$ & $6.1(3.2)\times10^{-5}$ & $5.8(3.4)\times10^{-5}$\\
		$f_\pi^2/f_d^2$ & 0.1023(35) &  0.1089(41)\\
		$m_d^2 / f_d^2$ & 1.94(65) &  2.24(80)\\
		\hline
		$l_a$ & --- & 0.78(27)\\
		\hline\hline
		$\chi^2/\text{dof}$ & 21.3/19 & 10.3/18\\
		AIC & 33.3 & 24.3\\
		\hline
	\end{tabular}
	\caption{Central values and uncertainties for LO (six-parameter) and NLO (seven-parameter) fits to the dEFT. Lattice data for $M_{\pi}$, $M_d$, $F_\pi$, $F_S$ and the $I=2$ scattering length has been incorporated into this fit, for 5 different vales of the underlying fermion mass $m$, corresponding to 25 data points. All dimensionful quantities are presented in units of the lattice spacing.}
	\label{Tab:deft}
\end{table}

We construct a chi-square function and minimize it with respect
to the six model parameters to find $\chi^2_{\text{min}}/N_{\text{dof}}=1.12$, indicating that lowest-order dEFT already provides a remarkably good fit to this expanded dataset.

At the chi-square minimum, the model parameters take the best-fit values shown in the LO column of Table~\ref{Tab:deft}. The fit parameters are consistent with those from earlier dEFT studies of the $N_f=8$ theory~\cite{Appelquist:2017wcg,Appelquist:2019lgk,Fodor:2019vmw}, which do not use $F_S$ or scattering data. There are also differences with respect to the fit results of Ref.~\cite{LatticeStrongDynamicsLSD:2021gmp}, which we attribute to an improved analysis of systematic errors. Table~\ref{Tab:deft} also reports the AIC value (Eq.~(\ref{eq:aic})).

\begin{figure}
	\centering
	\includegraphics[width=0.45\columnwidth]{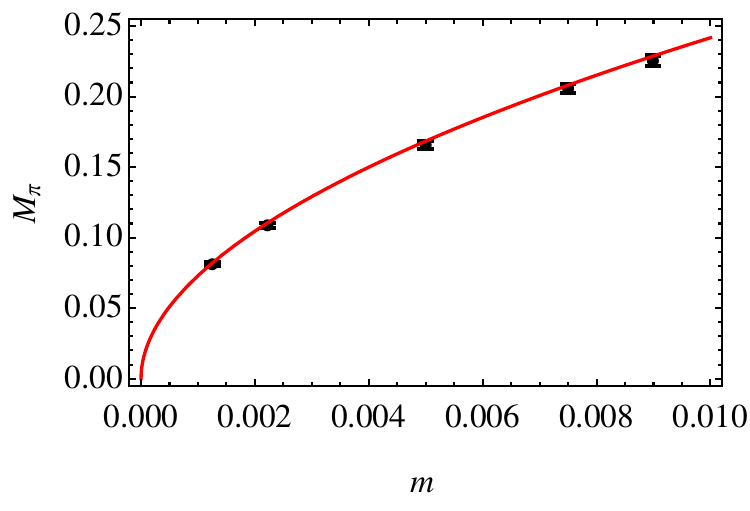}\quad
	\includegraphics[width=0.45\columnwidth]{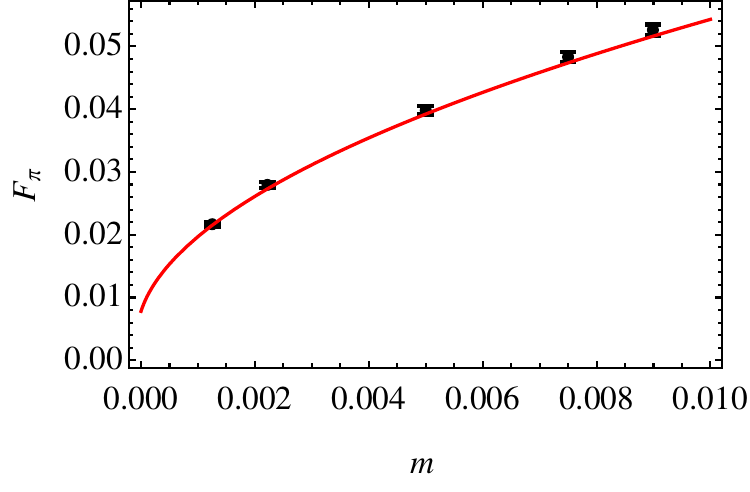}
	\includegraphics[width=0.45\columnwidth]{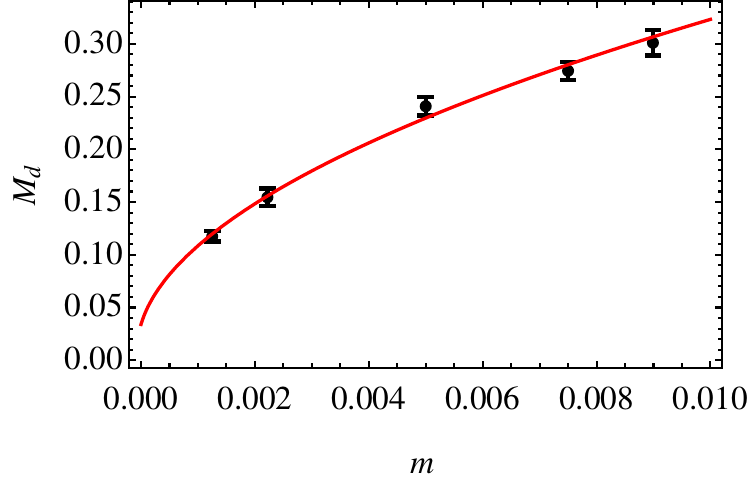}\quad
	\includegraphics[width=0.45\columnwidth]{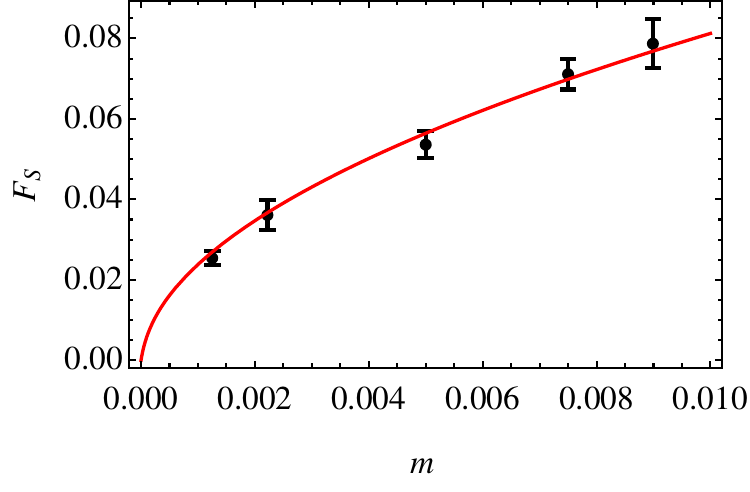}
	\caption{Lattice data for $M_\pi$, $M_d$, $F_\pi$, $F_S$ plotted against the underlying fermion mass $m$. The lines represent the dependence of these quantities on fermion mass predicted by LO dEFT, with the six EFT parameters set to their central values indicated in Table~\ref{Tab:deft}.}
	\label{fig:mandf}
\end{figure}

%The fit results in Table.~\ref{Tab:deft} indicate that the ratio of the physical dilaton mass to the pNGB decay constant extrapolates to $m^2_d/f^2_\pi\sim 20$ as the chiral limit $m^2_\pi\rightarrow0$ is approached. By contrast, in real world QCD, the analogous scalar $f_0(500)$ state has a larger mass in units of the pion decay constant: $m^2_{f0}/f^2_\pi\sim 30$.

In Fig.~\ref{fig:mandf}, we plot lattice data for the masses and decay constants for several values of the fermion mass. The plots confirm that dEFT accurately describes this lattice data, and show predictions for these quantities at smaller values of the fermion mass, which can be checked in the future. In the $m\rightarrow 0$ limit, we expect $F_{\pi}$ and $M_d$ to extrapolate to positive nonzero values.

Since all the quantities appearing on the  right side of Eq.~(\ref{eq:dilfspred}) were well determined in earlier dEFT fits~\cite{Appelquist:2017wcg,Appelquist:2019lgk,Fodor:2019vmw,Golterman:2020tdq,LatticeStrongDynamicsLSD:2021gmp}, Eq.~(\ref{eq:dilfspred}) can be viewed as a dEFT prediction for $F_S$. The new lattice measurements of $F_S$ align nicely with this prediction, providing new evidence for the dilaton interpretation of the light scalar.

\begin{figure}
	\centering
\includegraphics[width=0.7\columnwidth,trim= 0.5in 0in 0in 0in]{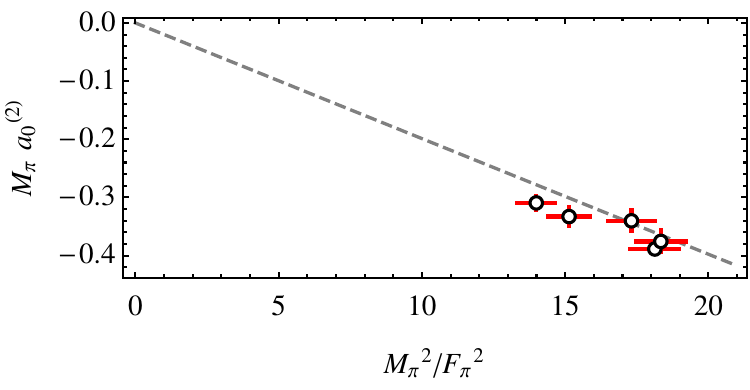}
\caption{The points represent lattice determinations of the pNGB scattering length in the $I=2$ channel plotted against $M^2_\pi/F^2_\pi$. The gray dashed line represents the LO dEFT prediction.}
\label{fig:scat}
\end{figure}

In Fig.~\ref{fig:scat}, we plot the s--wave pNGB scattering length in the
$I=2$ isospin channel against $M_{\pi}^2 / F_{\pi}^2$.
We assume here that the scattering length is well approximated
by $\left(k \cot \delta\right)^{-1}$ measured on the largest available lattice volume. The dEFT prediction, represented by the gray dashed line in Fig.~\ref{fig:scat}, is determined by inputting the LO central values for the $6$ dEFT parameters from Table~\ref{Tab:deft} into Eqs.~(\ref{eq:min})--(\ref{eq:dilatonslength}). Since $y-2 \ll 1$ and $f_{\pi}^2 / f_d^2 \ll 1$, the second term in Eq.~(\ref{eq:dilatonslength}) is negligible compared with the first, and the gray dashed line is approximately straight with a gradient of $ -1/16 \pi$. The dashed line passes close to all the points in Fig.~\ref{fig:scat}, indicating agreement between leading--order dEFT and the scattering data, similarly to what was shown in
Ref.~\cite{LatticeStrongDynamicsLSD:2021gmp}.

Finally we note that including data for $F_S$ in the global fit reduced the uncertainty in $\Delta$ considerably compared to previous fits, such that the value $\Delta\rightarrow4$ falls outside the $1\sigma$ range, which runs from 2.5 to 3.5.

% ------------------------------------------------------------------
\section{Mass--Deformed Conformal Theory}
\label{sec:mconf}

While dEFT provides a good fit to the lattice data, indicating that the $N_f = 8$ theory is outside the conformal window, it is important to test the alternate possibility that the theory is inside the conformal window, described in the infrared by mass-deformed conformal symmetry~\cite{DelDebbio:2010ze,DelDebbio:2010jy}. The analysis is simplest assuming that the gauge coupling can be approximated by its value at the infrared fixed point. This approach was employed in Ref.~\cite{Appelquist:2011dp} to describe the $SU(3)$ gauge theory with $N_f = 12$ Dirac fermions.

We proceed in the same manner, with the caveat that in the present case, the gauge coupling in the interval between the UV cutoff $\Lambda$ and the induced confinement scale might not have evolved to be sufficiently close to its fixed point value \cite{Cheng:2013xha}. With the assumption that it has, at scales $\mu\ll\Lambda$, the running fermion mass can be approximated by $m(\mu)=m(\Lambda/\mu)^{\gamma^*}$. Here $\gamma^{*}$ is the mass anomalous dimension at the fixed point, and $m=m(\Lambda)$ is the input fermion mass defined at the UV cutoff, which we identify with the inverse lattice spacing.

A simple scaling argument then leads to the result that for each physical state $X$, the mass $M_X$ is given to leading order (LO) by
\begin{align}
M_X = C_X m^{[1/(1+\gamma^*)]}\,,
\label{eq:mdcftm}
\end{align}
With all masses in units of $\Lambda$, the dimensionless coefficients $C_X$ are expected to be of order unity. An analogous expression applies to $F_{\pi}$ and $F_S$.

We also expect other quantities
with dimensions of mass to scale with $m$ in a similar
way. For example, the inverse scattering length should scale
as
\begin{align}
	1/a^{(2)}_0 = C_a\,m^{[1/(1+\gamma^*)]}\,.
	\label{eq:mdcfta}
\end{align}

We use these LO expressions to fit the lattice data for the masses and decay constants of the pseudoscalars $\pi$ and scalar $S$, as well as the inverse s--wave pseudoscalar scattering length in the $I = 2$ isospin channel, $1/a_0^2$.

Our leading-order fit uses six parameters, the same number as with the leading-order dEFT fit. The data set is also the same, employing five fermion mass values.  The result is shown in Table~\ref{tab:mdCFT}. A comparison with the leading-order dEFT fit of Table~\ref{Tab:deft} shows that the CFT fit has an AIC of $60.1$ compared to the dEFT AIC of $33.3$. The dEFT fit is thus favored over the CFT fit.

\begin{table}
	\centering
	\begin{tabular}{|c||c|c|c|}
		\hline
		Parameter & LO & NLO 1 & NLO 2 \\
		\hline\hline
		$C_{M_{\pi}}$ & 2.121(78) & 1.56(11)  &  1.57(12)  \\
		$C_{F_{\pi}}$ & 0.522(19) & 0.445(21) & 0.448(23) \\
		$C_{M_d}$ & 2.97(12)  & 2.53(12)  & 2.55(13)  \\
		$C_{F_S}$ & 0.706(33)  & 0.599(33) & 0.459(63) \\
		$C_{a}$  &  -5.88(22)  & -5.05(24) & -5.86(53) \\
		$\gamma^*$ & 1.073(28)  & 1.207(41) & 1.200(44) \\
		\hline
		$D_{M_{\pi}}$ & --- & 4.80(87)  & 4.71(90)  \\
		$D_{F_S}$ & --- & ---       & 2.77(98)  \\
		$D_{a}$  & --- & ---       & 12.9(5.8) \\
		\hline\hline
		$\chi^2/\text{dof}$
		& 48.1/19  & 20.9/18   & 6.90/16 \\
		AIC        & 60.1   & 34.9      & 24.9    \\
		\hline
	\end{tabular}
	\caption{\label{tab:mdCFT}Central values and uncertainties for the six parameter leading-order mass-deformed conformal fit are presented in the column labeled LO. Next-to-leading order fits are shown in the next two columns. All dimensionful quantities are expressed in units of the lattice spacing.}
\end{table}

%--------------------------------------------------------------------------
\section{Next-to-Leading Order}
\label{sec:nlo}

The infrared properties of the $N_f=8$ theory can be tested by the inclusion of next-to-leading order (NLO) terms in each fit.

\textbf{dEFT -} Since the specific forms of the NLO corrections to the dEFT fit equations are not all known, we adopt a more limited approach to account for their effects. It is evident from Fig.~\ref{fig:scat} that the greatest tension in the LO dEFT fit arises from the lattice data for the $I=2$ scattering length. There are NLO corrections to this quantity involving higher powers of $M_{\pi}^{2}/(4\pi F_{\pi})^2$ together with chiral logarithms. As a first NLO step, we modify only the LO expression, Eq.~(\ref{eq:dilatonslength}), by the addition of a single term of relative order $M_{\pi}^{2}/(4 \pi F_{\pi})^2$, disregarding the possible presence of a chiral logarithm. Thus we employ the expression
\begin{align}
	M_\pi a^{(2)}_{0} = \frac{-M^2_\pi}{16\pi F^2_\pi}\left(1-(y-2)^2\frac{f^2_\pi}{f^2_d}\frac{M^2_\pi}{M^2_d}+\frac{l_a M^2_\pi}{(4\pi F_\pi)^2}\right)\,,
	\label{eq:nloscat}
\end{align}
where $l_a$ is a new, seventh fit parameter. The result is shown in the NLO column of Table~\ref{Tab:deft}. The six parameters of the LO fit are changed only slightly, the parameter $l_a$ is of order unity, the $\chi^{2}$/dof falls by roughly a factor of two, and the AIC improves from $33.3$ to $24.3$. The NLO correction to the scattering length is less than 10\% throughout the range of lattice data. We also performed fits that included similar NLO corrections to the other measured quantities. However, the AIC for these fits was always higher than for the NLO fit shown in Table~\ref{Tab:deft}.

\textbf{CFT -} Maintaining the assumption that the gauge coupling takes its infrared fixed point value, higher order corrections to the fit equations (\ref{eq:mdcftm}) - (\ref{eq:mdcfta}) enter additively as polynomials in $m$. The leading (NLO) correction in each case is linear
in $m$, arising from physics near the UV cutoff. Each can be expected to be small compared to the LO term
provided only that $\gamma^* > 0$. For example, we can modify Eq.~(\ref{eq:mdcftm}) to read
\begin{align}
	M_X = C_X m^{[1/(1+\gamma^*)]}+D_X\,m\,,
	\label{eq:mdcftmnlo}
\end{align}
where $D_X$ is a new fit parameter.

If we modify only a single fit equation in this way, the one that leads to the smallest AIC is the equation for $M_{\pi}$. The result is shown in the column labeled NLO 1 in Table~\ref{tab:mdCFT}. The six parameters of the LO fit are changed by small amounts. While both the $\chi^2$/dof and AIC improve, they remain poor relative to those of the NLO fit for the dEFT. 

Further reduction in the $\chi^{2}$/dof and the AIC for the mass-deformed CFT fit can be achieved by including the analogous NLO modification of other fit equations. The smallest AIC comes from adding NLO modifications to only the equations for $F_S$ and the $I=2$ scattering length. The result is shown in the column of Table~\ref{tab:mdCFT} labeled NLO 2, where the AIC is reduced to $24.9$, which is comparable to the AIC of the NLO dEFT fit.

But the column labeled NLO 2 also reveals that some of the NLO corrections required to fit the data are not much smaller than the leading order terms in this fit. In particular, the NLO correction to $F_S$ grows to 46\% of the LO term size at the top of the fermion mass range, indicating that the expansion in powers of $m$ is not under good theoretical control.

% ------------------------------------------------------------------
\section{Discussion}
\label{sec:disc}

We analyzed the latest data for an $SU(3)$ gauge theory with $N_f=8$ Dirac fermions in the fundamental representation, focusing on the masses and decay constants of the lightest composite states. These comprise a set of pseudoscalars and one additional scalar. 

We first adopted the view that the gauge theory is just outside the conformal window, leading to the interpretation of the pseudoscalars as approximate Nambu-Goldstone bosons. We interpreted the scalar as an approximate dilaton, associated with the spontaneous breaking of conformal symmetry. We employed a dilaton effective field theory (dEFT) to fit the lattice data, newly expanded to include results for a scalar decay constant and the scattering length in the $I=2$ channel. At leading order, the result of the dEFT fit is summarized in Table~\ref{Tab:deft}. The goodness of the fit is notable. 

We then considered the possibility that the gauge theory is inside the conformal window. In this case, the fermion mass $m$ provides a small deformation of infrared conformal symmetry. Simple scaling laws were employed to describe the masses and decay constants of the light states as a function of $m$. The result of a leading-order (LO) fit, employing six parameters as in the case of the LO dEFT fit, is summarized in Table~\ref{tab:mdCFT}. The $\chi^{2}$/dof and AIC are both larger than those of the dEFT, indicating a lower goodness-of-fit.

We also examined the quality of fits with the inclusion of a next-to-leading-order (NLO) term. For both the dEFT and the mass-deformed CFT, the goodness-of-fit was improved but the dEFT fit remained superior. The addition of extra NLO parameters in the dEFT fit led to no further improvement, while it did improve the mass-deformed-CFT fit. But to bring this goodness-of-fit to the level of the dEFT fit led to uncomfortably large NLO corrections to the LO term.

Finally we note again that both the dEFT and mass-deformed-CFT fits involved the assumption that the underlying gauge theory is nearly conformal, with the gauge coupling approximated by its fixed-point value for energies above the induced confinement scale. The lesser quality of the mass-deformed-CFT fit could be a consequence of this assumption not being satisfied \cite{Cheng:2013xha}. Lattice simulations at stronger couplings might then be required. For the dEFT fit, the assumption of near-conformality was critical for the identification of the scalar as an approximate dilaton. It will be interesting to test whether dEFT describes other systems which could be approximately conformal \cite{LatticeStrongDynamics:2020uwo}. The success of the dEFT fit presented here, incorporating newly expanded and refined lattice data, is quite striking.

% ------------------------------------------------------------------
\begin{acknowledgments}
  R.C.B.~and C.R.~acknowledge United States Department of Energy (DOE) Award No.~{DE-SC0015845}.
  K.C.~acknowledges support from the DOE through the Computational Sciences Graduate Fellowship (DOE CSGF) through grant No.~{DE-SC0019323}, and also from the P.E.O Scholar award.
  G.T.F.~acknowledges support from DOE Award No.~{DE-SC0019061}.
  A.D.G.~is supported by SNSF grant No.~{200021\_17576}.
  A.H.~and E.T.N.~acknowledge support by DOE Award No.~{DE-SC0010005}.
  J.I.~acknowledges support from ERC grant No.~{101039756}.
  D.S.~was supported by UK Research and Innovation Future Leader Fellowship No.~{MR/S015418/1} and STFC grant {ST/T000988/1}.
  P.V.~acknowledges the support of the DOE under contract No.~{DE-AC52-07NA27344} (Lawrence Livermore National Laboratory, LLNL).

  We thank the LLNL Multiprogrammatic and Institutional Computing program for Grand Challenge supercomputing allocations. We also thank Argonne Leadership Computing Facility (ALCF) for allocations through the INCITE program. ALCF is supported by DOE contract No.~{DE-AC02-06CH11357}. Computations for this work were carried out in part on facilities of the USQCD Collaboration, which are funded by the Office of Science of the DOE, and on Boston University computers at the MGHPCC, in part funded by the National Science Foundation (award No.~{OCI-1229059}). This research utilized the NVIDIA GPU accelerated Summit supercomputer at Oak Ridge Leadership Computing Facility at the Oak Ridge National Laboratory, which is supported by the Office of Science of the U.S. Department of Energy under Contract No. DE- AC05-00OR22725.
\end{acknowledgments}

% --------------------------------------------------------------------

% ------------------------------------------------------------------
\raggedright
\bibliography{nf8short}
\end{document}